\documentstyle[11pt,aaspp4,flushrt]{article}
\input psfig

\newcommand\PA{\mbox{PA}}
\newcommand\th{\theta}
\newcommand\k{\kappa}
\newcommand\g{\gamma}
\newcommand\tg{\th_\g}
\newcommand\s{\sigma}
\newcommand\ts{\th_\s}
\newcommand\dd{\delta}
\newcommand\td{\th_\dd}
\newcommand\al{\alpha}
\newcommand\kms{\hbox{km s$^{-1}$}}
\newcommand\Hunits{\hbox{km s$^{-1}$ Mpc$^{-1}$}}
\newcommand\Scrit{\Sigma_{crit}}
\newcommand\Sc{\Sigma_{clus}}
\newcommand\kc{\kappa_{clus}}
\newcommand\gc{\gamma_{clus}}
\newcommand\tc{\th_{clus}}
\newcommand\geff{\gamma_{\rm eff}}
\newcommand\xx{{\vec x}}
\newcommand\uu{{\vec u}}

\newcommand\cbk{\chi_{bk}^2}
\newcommand\cf{\chi_{flux}^2}
\newcommand\ca{\chi_{arc}^2}

\newcommand\avg[1]{\langle{#1}\rangle}

\newcommand\refeq[1]{eq.~(\ref{eq:#1})}

\newcommand\map[1]{}

\slugcomment{Submitted to ApJ}

\lefthead{Keeton et al.}
\righthead{Q~0957+561 Host Galaxy}

\begin{document}

\title{The Host Galaxy of the Lensed Quasar Q~0957+561\footnote{Based on
Observations made with the NASA/ESA Hubble Space Telescope, obtained at
the Space Telescope Science Institute, which is operated by AURA, Inc.,
under NASA contract NAS 5-26555.}}

\author{C. R. Keeton$^{(a)}$,}
\author{E. E. Falco$^{(b)}$, C. D. Impey$^{(a)}$, C. S. Kochanek$^{(b)}$, J. Leh\'ar$^{(b)}$, B. A. McLeod$^{(b)}$,}
\author{H.-W. Rix$^{(c)}$, J. A. Mu\~noz$^{(b)}$, and C. Y. Peng$^{(a)}$}

\affil{$^{(a)}$Steward Observatory, University of Arizona, Tucson, AZ 85721}
\affil{email: ckeeton, cimpey, cyp@as.arizona.edu}
\affil{$^{(b)}$Harvard-Smithsonian Center for Astrophysics, 60 Garden St.,
  Cambridge, MA 02138}
\affil{email: efalco, ckochanek, jlehar, bmcleod, jmunoz@cfa.harvard.edu}
\affil{$^{(c)}$Max-Planck-Institut f\"ur Astronomie, Koenigsstuhl 17,
  D-69117 Heidelberg, Germany}
\affil{email: rix@mpia-hd.mpg.de}

\begin{abstract}
Infrared images of the Q~0957+561 gravitational lens obtained with
the Hubble Space Telescope show two large ($\sim$5\arcsec) lensed
images of the $z_s=1.41$ quasar host galaxy. Parts of the host
galaxy are doubly-imaged like the quasar, while other parts are
quadruply-imaged. The distortions of the host galaxy offer the best
probe yet of the global structure of the lensing potential, which
is essential for determining the Hubble constant from the measured
time delay. The distortions are inconsistent with the predictions
of previously published lens models, which invalidates those models
and their implications for $H_0$. New models show that the
distortions finally break the long-standing degeneracy between the
shape of the lens galaxy and the tidal shear contributed by the
cluster containing the lens galaxy. The shape of the lens galaxy's
mass distribution must be remarkably similar to the shape of its
luminosity distribution, and most models that produce reasonable
values for the Hubble constant roughly match the observed
ellipticity gradient and isophote twist of the lens galaxy. Also,
the cluster must be non-spherical and produce a relatively small
tidal shear. Although there are still degeneracies in the lens
models that lead to a 25\% uncertainty in the derived value of the
Hubble constant, there are also strong prospects for new
observations to further improve the constraints and reduce the
uncertainties.
\end{abstract}

\section{Introduction}

Gravitational lenses offer an attractive independent method of
determining the Hubble constant ($H_0$) on cosmological scales
without the systematic difficulties associated with the local
distance ladder (Refsdal 1964, 1966). Six of the more than 50 known
gravitational lenses now have time delays that can be used to
estimate $H_0$: B~0218+357 (Biggs et al.\ 1999); Q~0957+561
(Schild \& Thomson 1995; Kundi\'c et al.\ 1997; Haarsma et al.\
1999); PG~1115+080 (Schechter et al.\ 1997; Barkana 1997);
B~1600+434 (Hjorth et al.\ 1999); B~1608+656 (Fassnacht et al.\
1999); and PKS~1830$-$211 (Lovell et al.\ 1998); preliminary
$H_0$ estimates for five of these systems have been compiled
by Koopmans \& Fassnacht (1999). Once a time delay is accurately
determined, the uncertainties in the derived value of the Hubble
constant are due almost entirely to the systematic uncertainties
in the model for the lensing potential, plus ``cosmic variance''
due to the effects of weak density inhomogeneities along the line
of sight (Seljak 1994; Barkana 1996).

Only two of the time delay lenses have been modeled in sufficient
detail to fully understand the systematic uncertainties created by
their geometries: Q~0957+561 (Falco, Gorenstein \& Shapiro 1985,
1991; Kochanek 1991; Bernstein, Tyson \& Kochanek 1993; Grogin \&
Narayan 1996; Chartas et al.\ 1998; Barkana et al.\ 1999; Bernstein
\& Fischer 1999; Chae 1999; Romanowsky \& Kochanek 1999), and
PG~1115+080 (Schechter et al.\ 1997; Courbin et al.\ 1997; Keeton
\& Kochanek 1997; Saha \& Williams 1997; Impey et al.\ 1998).  In
Q~0957+561, the lens consists of a brightest cluster galaxy and its
parent cluster, and the value of the Hubble constant depends on the
mass balance between the two components. Judgments about the
ability of models to determine the correct mass balance vary
significantly: optimistic estimates yield $H_0 = 61_{-15}^{+13}$
\Hunits at 95\% confidence (Grogin \& Narayan 1996, using the
stellar dynamical models of Romanowsky \& Kochanek 1999); while
more pessimistic estimates yield $H_0 = 77_{-24}^{+29}$ (Kochanek
1991; Bernstein et al.\ 1993; Bernstein \& Fischer 1999).
In PG~1115+080, the lens is one of the brighter galaxies in a small
group, but the four-image geometry easily determines the relative
roles of the primary lens galaxy and the group. Instead, the value
of the Hubble constant depends on the assumed radial mass
distribution of the primary lens galaxy. Galaxy models with dark
matter and mass distributions consistent with the best estimates
for early-type galaxies lead to low values for the Hubble constant
($H_0 = 44\pm4$ \Hunits; Impey et al.\ 1998).

Impey et al.\ (1998) discovered an Einstein ring image of the
quasar host galaxy in the PG~1115+080 lens, and noted that the
geometry of the ring could be used to break the degeneracy in the
lens models. Here we report the discovery of the host galaxy in the
Q~0957+561 lens and discuss the implications for lens models and
the Hubble constant. In \S 2 we describe our observations. In \S 3
we discuss the data and methods used to model the system,
summarizing previous work and introducing our new data and
techniques. In \S 4 we examine the arcs predicted by previously
published lens models and demonstrate that they fail to match the
observed arcs. In \S 5 we present new models that match all data,
including the arcs, and discuss their implications for the
properties of the lens galaxy and cluster and for $H_0$. In \S 6
we summarize our results and discuss prospects for further
improving the constraints on this lens system and on $H_0$.

\section{Observations}

We observed Q~0957+561 with the Hubble Space Telescope as part of
the CfA--Arizona Space Telescope Lens Survey (CASTLES; Leh\'ar et
al.\ 1999; Falco et al.\ 1999).  Using the NIC2 camera, we obtained
a 2800-second F160W (H band) image of Q~0957+561 divided into four
dithered exposures. A log of the observations is presented in
Table 1. We reduced the images using {\tt nicred}, a custom
reduction package developed for CASTLES (Leh\'ar et al.\ 1999).
We also reanalyzed archival WFPC2 images, including a 32200-second
F555W (V band) exposure and a 2620-second F814W (I band) exposure
(Bernstein et al.\ 1997). In the optical images, Q~0957+561 fell
in chip WF3 of the WFPC2 camera (with pixel size $\sim$0\farcs1,
compared to $\sim$0\farcs076 for NIC2). Because the images A and
B of Q~0957+561 were saturated in the WFPC2 images, we derived
our astrometry of A, B and the lens galaxy G1 exclusively from
our unsaturated NIC2 data.

Figure 1a shows our combined H band image, which prominently shows
the lens galaxy and the two quasar images. The quasar host galaxy
is visible as a faint arc next to the quasar A image. Table 2
summarizes a photometric model consisting of point sources for the
two quasar images and elliptical de Vaucouleurs models for the main
lens galaxy (G1) and a neighboring galaxy (G2, not seen in Figure
1).\footnote{Young et al.\ (1981) tabulated objects in the field
of Q~0957+561 and applied the labels G1--G5 to five of the bright
galaxies.  We labeled only objects close to the lens galaxy, so
other than G1 our labels do not match those of Young et al.  Our
object G2 appears as object \#97 in their Table 1.  Our objects
G3 and G4 do not appear in their list.}  The lens galaxy is known
to have a small ellipticity gradient and isophote twist (Bernstein
et al.\ 1997), but we used only a simple elliptical model.  Figure
1b shows the residuals after subtracting the PSF-convolved
photometric model from the original image, and Figure 1c shows
these residuals convolved with a $\sim$0\farcs076 (1 pixel) FWHM
Gaussian to enhance the visibility of low surface brightness
features. The residual image shows the host galaxy image
near quasar A to be an arc distorted tangentially relative to the
lens galaxy. It also reveals an extended asymmetric arc near the B
quasar image. If the residuals near quasar B were created by the
error of fitting a simple elliptical surface brightness model to a
galaxy with a radially varying ellipticity and orientation, we
would expect them to have reflection symmetry through the center
of the galaxy. Because the arc near quasar B lacks such symmetry,
we conclude that it is the lensed counterpart of the arc near
quasar A, and that both are images of the quasar host galaxy.

The residual H band image also shows two additional sources, one east
of quasar A and another just west of quasar B (labeled G3 and G4 in
Figure 1c). Their positions and photometric properties are given in
Table 2. These objects correspond in position and V magnitude to
two of the faint sources seen in the deep V band image and labeled
Blobs 7 and 1 (respectively) by Bernstein et al.\ (1997). Their
colors (especially V$-$I) are similar to those of G1 and G2 and
suggest that they are probably faint cluster members.  Whether or
not they are associated with the cluster, it is clear that they are
not multiply imaged and are not associated with the quasar source.
The fact that G4 is not part of the host galaxy arc will be
important to remember when examining the arc structure predicted
by lens models (see \S 3.3).

The lensed host galaxy is not detected at significant levels in
the shallow I band image, although a visual examination suggests
that low-level residuals are present and should be detected at a
significant level with a longer integration. The deep V band image
does not have large extended arcs, but it does show seven faint
``blobs'' and a thin arclet (Bernstein et al.\ 1997).  As noted
above, at least two of the blobs are probably faint cluster
galaxies.  However, Blobs 2 and 3 and the arclet have been
identified as possible lensed features, and lens models support
this interpretation (Barkana et al.\ 1999; Bernstein \& Fischer
1999; Chae 1999).  Since the V band corresponds to rest-frame UV
at the quasar redshift $z_s=1.41$, these features probably
correspond to discrete star forming regions in the host galaxy.
We follow previous models and include Blobs 2 and 3 plus two
``Knots'' in the arclet as model constraints (see \S 3.2).

We estimated the fluxes of the host galaxy images using the IRAF
task {\tt polyphot}, which measures fluxes within polygonal
apertures. We traced polygonal apertures along the edges of the
distorted images of the host galaxy at the $3\sigma$ level above
the sky. We found that both lensed images of the host galaxy have H
band brightnesses of $\sim$18.4 mag, and a surface brightness of
$\sim$20.6 mag/arcsec$^2$ (both images have approximately the same
area). The main source of uncertainty for the brightness estimates
is the relatively large size of the residuals from the subtraction
of the quasar images, compared to the (low) brightness levels of
the arcs; we estimate the uncertainties to be $\sim$0.3 mag.

Given an acceptable lens model we can map the host galaxy images to
the source plane to obtain a map of the unlensed host galaxy (see
\S 3.3). In principle we could use the source maps to measure the
photometric properties of the host galaxy. However, in practice the
imperfect quasar subtraction corrupts the flux in the bright
central regions of the source and hinders the measurement. One
robust statement we can make is that part of the host galaxy is
doubly-imaged like the quasar, but part of it is quadruply-imaged
(see \S 4).  This accounts for the shape differences between the
A and B arcs: the A arc is a single distorted image of the host
galaxy, while the B arc is a composite of three images that
straddle the lensing critical line.

\section{Constraining models of Q~0957+561}

Q~0957+561 is the most thoroughly studied gravitational lens. The
system comprises a radio-loud quasar at redshift $z_s=1.41$ lensed
into two images by a brightest cluster galaxy and its parent
cluster at redshift $z_l=0.36$ (Walsh, Carswell \& Weymann 1979;
Young et al.\ 1980). There is a time delay of $417\pm3$ days
between the images (Schild \& Thomson 1995; Kundi\'c et al.\ 1997;
Haarsma et al.\ 1999). VLBI observations have resolved each image
into a core and $\sim$80 milli-arcsecond jet (Garrett et al.\
1994). Deep optical images have uncovered faint ``Blobs'' and
``Knots'' that are probably lensed image pairs of star forming
regions in the quasar host galaxy (Bernstein et al.\ 1997). Our
observations have revealed infrared arcs representing distorted
images of the quasar host galaxy. In this section we discuss how
to use these and other data to constrain models of the system and
values for the Hubble constant $H_0$. In \S 3.1 we review the
theory of lens modeling and discuss important model degeneracies.
In \S\S 3.2 and 3.3 we discuss previous and new observational
constraints on models. In \S\S 3.4 and 3.5 we describe classes
of models applied to the lens galaxy and cluster in Q~0957+561,
summarizing previous classes and introducing a new one. Finally,
in \S 3.6 we discuss how we apply the observational constraints
to our new class of models.

\subsection{Basic lens theory: model degeneracies}

In a multiply-imaged gravitational lens system, the light from a
distant source is deflected by the gravitational potential of
foreground objects so that we observe multiple images of the
source. The lensing potential is usually dominated by a single
galaxy, although there may be non-negligible perturbations from
other objects nearby.\footnote{There may also be a contribution
to the potential from density fluctuations along the line of sight,
but it is usually small compared to the contribution from objects
at the same redshift as the main lens galaxy (Seljak 1994; Barkana
1996; Keeton, Kochanek \& Seljak 1997). In a few cases there are
two lens galaxies enclosed by the lensed images (e.g.\ Jackson,
Nair \& Browne 1997; Koopmans et al.\ 1999), but this complication
is not an issue in Q~0957+561.} Basic lens theory is presented in
the book by Schneider, Ehlers \& Falco (1992), and we quote the
relevant results here.  The lensing potential $\phi$ is determined
by the two-dimensional Poisson equation $\nabla^2\phi = 2\k$,
where $\k = \Sigma/\Scrit$ is the surface mass density in units
of the critical surface density for lensing (in angular units),
\begin{equation}
  \Scrit = {c^2 \over 4\pi G}\,{D_{ol} D_{os} \over D_{ls}}\ ,
  \label{eq:Scrit}
\end{equation}
where $D_{ol}$ and $D_{os}$ are angular diameter distances from the
observer to the lens and source, respectively, and $D_{ls}$ is the
angular diameter distance from the lens to the source. The lensing
potential deflects a light ray so that the angular position $\uu$
of the source on the sky and the angular position $\xx$ of an image
are related by the lens equation,
\begin{equation}
  \uu = \xx - {\vec\nabla}\phi(\xx)\,. \label{eq:lens}
\end{equation}
There is an image corresponding to each solution $\xx_i$ of this
equation.  Lensing introduces a time delay between the ray paths
of two images of the same source. The time delay between images
at positions $\xx_i$ and $\xx_j$ is
\begin{equation}
  \Delta t_{ij} = {1+z_l \over c}\,{D_{ol} D_{os} \over D_{ls}}
    \left[ {1\over2}\Bigl( |\xx_i-\uu|^2 - |\xx_j-\uu|^2 \Bigr) -
    \Bigl( \phi(\xx_i)-\phi(\xx_j) \Bigr) \right] ,
    \label{eq:tdel}
\end{equation}
where $z_l$ is the redshift of the lens. This equation is the basis
of attempts to use lensing to determine the Hubble constant $H_0$.
By measuring light curves of images one can determine the time
delay $\Delta t_{ij}$. A lens model gives the term in square
brackets in \refeq{tdel}. The combination of distances is $\propto
H_0^{-1}$ and only weakly dependent on other cosmological
parameters.

A lens model consists of a description of the lensing potential
$\phi$. The observed images provide two principle constraints on
such a model. First, roughly speaking an image at projected
distance $R$ from the main lens galaxy measures the enclosed mass
$M(R)$. In Q~0957+561 and many 2-image gravitational lenses, the
two images lie at different distances $R_1 \ne R_2$, so they
measure two masses $M(R_1)$ and $M(R_2)$ and hence constrain the
mass profile (e.g.\ Grogin \& Narayan 1996). Second, with at least
three well-determined positions (two images and the lens galaxy, or
four images), the images determine the quadrupole moment of the net
potential. However, there are four properties of the mass
distribution that must be determined for a complete description of
the model: the mass profile of the main lens galaxy; the shape
(ellipticity and orientation) of the main lens galaxy; the shear
from the the gravitational tidal field induced by objects near the
lens galaxy; and the amount of gravitational focusing (or
``convergence'') contributed by the environment of the lens galaxy.

With more unknown quantities than constraints, there are two common
degeneracies in the lens models. First, to lowest order the
convergence $\k$ contributed by the environment cannot be
determined by lens models, which leads to the so-called ``mass
sheet degeneracy'' (Falco et al.\ 1985). If a lensing potential
$\phi(\xx)$ fits the observed data, then any potential
\begin{equation}
  \phi'(\xx) = {1\over2}\,\k\,|\xx|^2 + (1-\k)\,\phi(x)
  \label{eq:phik}
\end{equation}
will fit the data equally well. The $\k$ term is equivalent to the
potential from a uniform mass sheet with surface density $\Sigma =
\k\,\Scrit$. The only potentially observable difference between the
lens models represented by $\phi$ and $\phi'$ is in the predicted
time delays,
\begin{equation}
  \Delta t_{model}' = (1-\k)\,\Delta t_{model}\,. \label{eq:dtk}
\end{equation}
However, this effect is not observable if we want to use
gravitational lensing to determine the Hubble constant because it
simply translates into a scaling of the inferred value for $H_0$.
If $H_0$ and $H_0'$ are the values inferred from the two lens
models, then
\begin{equation}
  H_0' = (1-\k)\,H_0\,. \label{eq:hk}
\end{equation}
This analysis has used only the lowest order term of the contribution
from the environment, but the higher order terms cannot necessarily
eliminate the degeneracy (e.g.\ Chae 1999).  The mass sheet degeneracy
is important for Q~0957+561 because the parent cluster of the lens
galaxy contributes a significant convergence. The only way to break
this degeneracy is to obtain an independent mass constraint to
determine the relative contributions of the main lens galaxy and
the environment to the mass enclosed by the images (see \S 3.2).

The second important degeneracy is between the shape of the main
lens galaxy and the shear from the gravitational tidal field of
objects in the environment of the lens galaxy. Because the images
determine only the quadrupole moment of the total potential, there
can be a wide range of parameter space in which the lens galaxy
ellipticity and the external shear combine to produce the required
joint quadrupole (e.g.\ Keeton et al.\ 1997). Consider a 2-image
lens with image positions $\xx_A$ and $\xx_B$, a galaxy potential
$\phi_{gal}$, and an external shear with amplitude $\g$ and
direction $\tg$. Fitting the image positions exactly is equivalent
to solving the equation
\begin{equation}
  \xx_A - {\vec\nabla}\phi_{gal}(\xx_A) - \Gamma\cdot\xx_A =
  \xx_B - {\vec\nabla}\phi_{gal}(\xx_B) - \Gamma\cdot\xx_B
  \label{eq:shr}
\end{equation}
where the shear is described by the tensor
\begin{equation}
  \Gamma = \left[ \begin{array}{rr}
    \g \cos2\tg &  \g \sin2\tg \cr
    \g \sin2\tg & -\g \cos2\tg \cr
  \end{array} \right] .
\end{equation}
No matter what galaxy potential $\phi_{gal}$ is used, it is
straightforward to solve \refeq{shr} for the shear parameters
$\g$ and $\tg$. In other words, one can fit the image positions
in a 2-image lens to arbitrary precision for {\it any model lens
galaxy\/}, although in some cases the inferred shear $\g$ may be
unphysically strong.

Constraints from the images fluxes or additional images often
break this degeneracy, although the $1\sigma$ range of models may
still be large.  In previous models of Q~0957+561, the additional
constraints came primarily from the VLBI observations of the
quasar images, which show several discrete components with
$\sim$0.1 milli-arcsecond errorbars (Garrett et al.\ 1994;
Barkana et al.\ 1999).  It is easy to be misled about the true
structure of the potential, however, when using such strong
constraints (see Kochanek 1991; Bernstein et al.\ 1993; Bernstein
\& Fischer 1999).  Internal structure in the lens galaxy, such
as isophote twists or lumpiness in the mass distribution, may
affect the images at sub-milli-arcsecond scales (e.g.\ Mao \&
Schneider 1998).  Neglecting that structure forces the models
to adjust large-scale model components in order to fit small-scale
constraints, leading them to converge to a best-fit solution that
is well defined (i.e.\ with no apparent degeneracy) but incorrect.
In Q~0957+561, the observed ellipticity gradient and isophote
twist suggest that the galaxy does have important internal
structure (Bernstein \& Fischer 1997), and in \S 4 we show that
models neglecting this structure indeed converged to incorrect
solutions.

\subsection{Previous observational constraints}

In Q~0957+561 the high-resolution VLBI maps of the quasar images
provide strong position constraints, plus somewhat weaker
constraints on the relative magnification matrix between the images
(Garrett et al.\ 1994; Barkana et al.\ 1999). The optical Blobs
and Knots offer additional but weaker position constraints. The
Knots are particularly useful because they appear to represent
a pair of ``fold'' images and thus require the lensing critical
line to pass between them (see Bernstein \& Fischer 1999).
The constraints from these images leave two of the three model
degeneracies discussed in \S 3.1: the mass sheet degeneracy
due to the convergence provided by the cluster; and the
degeneracy between the lens galaxy shape and the cluster shear.

Several steps have been taken to break the mass sheet degeneracy
by determining the relative masses of the lens galaxy and cluster.
First, Rhee (1991), Falco et al.\ (1997), and Tonry \& Franx (1999)
measured the lens galaxy's central velocity dispersion to be
$288\pm9$ \kms. Romanowsky \& Kochanek (1999) used stellar dynamical
models to translate this into constraints on the lens galaxy mass
and concluded that $(80\pm12)\%$ of the image separation is
contributed by the galaxy (and the rest by the cluster convergence).
Second, Fischer et al.\ (1997) detected a weak lensing signal from
the cluster, which Bernstein \& Fischer (1999) translated into an
estimate of the mean surface density of all mass inside an aperture
of radius 30\arcsec\ centered on the lens galaxy,
\begin{equation}
  \avg{\kappa_{obs}}_{30\arcsec}
  = \avg{\Sigma}_{30\arcsec} / \Scrit
  = 0.26 \pm 0.08\,.
\end{equation}
Furthermore, Bernstein \& Fischer (1999) showed that the surface
mass density $\kc=\Sc/\Scrit$ of the cluster at the position of the
lens galaxy is given by
\begin{equation}
  1 - \kc = { 1 - \avg{\kappa_{obs}}_{30\arcsec} \over
    1 - \avg{\kappa_{mod}}_{30\arcsec} }\ , \label{eq:kc}
\end{equation}
where $\avg{\kappa_{mod}}$ is the mean mass density of a model lens
galaxy. In this way one can combine the weak lensing measurement
with a lens model to estimate $\kc$ (see \S 5.3). There have been
other attempts to measure the cluster mass, using the velocity
dispersion of the galaxies in the cluster ($715\pm130$ \kms;
Garrett, Walsh \& Carswell 1992; Angonin-Willaime, Soucail \&
Vanderriest 1994) or the X-rays from the cluster gas (Chartas et
al.\ 1998), but they are still limited by significant systematic
uncertainties.

Although measurements of the cluster surface density are important
(especially for determining $H_0$, see \S 5.4), they do not break
the degeneracy between the lens galaxy shape and the cluster shear.
Another approach is to constrain the shape of the lens galaxy's
mass distribution. The orientation of the lens galaxy's light
distribution should offer a guide to the orientation of its mass
distribution, because other lenses suggest that mass and light
are at least roughly aligned (typically within $\sim$10\arcdeg;
Keeton, Kochanek \& Falco 1998). An important detail, though, is
that the lens galaxy has a radially varying ellipticity and
orientation (Bernstein et al.\ 1997): the ellipticity increases
from $\sim$0.1 inside a radius of 1\arcsec\ to $\sim$0.4 outside
10\arcsec, and the PA varies from $\sim$40\arcdeg\ to
$\sim$60\arcdeg\ (albeit with large uncertainties). These
variations suggest that the galaxy's projected mass distribution
may not have simple elliptical symmetry, and they may well need to
be incorporated into lens models.

\subsection{New constraints from the host galaxy arcs}

The host galaxy arcs act as an extensive set of position
constraints that may break the degeneracy between the lens galaxy
shape and the cluster shear. Although the intrinsic source
structure is unknown, the simplicity of the lensing geometry means
that we can model the arcs self-consistently using a modified
version of the Ring Cycle algorithm of Kochanek et al.\ (1989).
Specifically, we know that there is a one-to-one mapping between
the A image of the host galaxy and the source, so for a given lens
model we can project the A arc back to the source plane to obtain
the correct source structure for that lens model.\footnote{A
similar projection fails for the B arc because in all reasonable
models this arc crosses the lensing critical line (see \S 4). Thus
the map from the B arc to the source is not one-to-one.} We can
then reproject that source onto the image plane to predict the
structure of the B arc. Since the arcs are very large compared to
the size of the PSF, there is no need to include the effects of the
PSF. For the arc map we use the smoothed residual image (Figure
1c).

This technique produces maps of the model arcs that can be compared
visually to the observed arcs (see Figures 2--4 below). However,
for modeling we want to quantify the differences, so we define a
$\chi^2$ term for the arcs using a pixel-by-pixel comparison of the
observed and model surface brightnesses,
\begin{equation}
  \ca = \sum_{i=1}^{N_{arc}} {(f_{obs,i}-f_{mod,i})^2 \over 2\sigma_{sky}^2}\ .
\end{equation}
The factor of 2 in the denominator enters because the source is
constructed from arc A, so it and the observed B arc each have
errorbars of $\sigma_{sky}$. To compute $\ca$ we mask the arc map
and use only the region near the B arc, omitting regions near the B
quasar and the core of the lens galaxy where imperfect subtraction
makes the residual flux unreliable. We also omit the region around
the object G4 where the flux is not due to the quasar host galaxy
(see \S 2). The masked region is shown in Figure 1c. To count the
number of arc constraints, we count the number of pixels inside the
mask for which either the observed arc or the model arc is more
than $2\sigma$ above the sky and take this to be the number of arc
constraints $N_{arc}$. Typically $N_{arc} \sim 10^4$.  Because
the source is fully determined from the A arc, we quote $\bar\ca
\equiv \ca/N_{arc}$ as an effective $\chi^2/\mbox{DOF}$ for the arc.
The fact that our best models produce $\bar\ca \simeq 1$ (see \S 5)
suggests that this is a reasonable counting of the constraints.

\subsection{Lens galaxy models}

Q~0957+561 has been modeled extensively; we summarize results from
the recent models by Grogin \& Narayan (1996), Barkana et al.\
(1999), Bernstein \& Fischer (1999), and Chae (1999) in Table 3.
(Technical details of comparing the models are discussed in
Appendix A.) The models fall into two main families based on how
they treat the main lens galaxy. First is the ``power law'' family,
in which the lens galaxy is modeled with a circular or elliptical
surface density with a softened power law profile: $\Sigma \propto
\left(s^2+m^2\right)^{\al/2-1}$, where $m$ is an ellipsoidal
coordinate, $s$ is a core radius, and $\al$ is the power law
exponent such that $M(R) \sim R^\al$ asymptotically. The best-fit
models typically have a small core radius and a power law $\al
\simeq 1.1$ corresponding to a profile slightly shallower than
isothermal. In the second family, called ``FGS'' models after
being introduced by Falco, Gorenstein \& Shapiro (1991), the lens
galaxy is treated as a circular or elliptical King model, plus
a central point mass to account for a mass deficit in the core
of the King model.

In almost all of the previous models the lens galaxy was assumed
to have a projected mass distribution with circular or elliptical
symmetry.  However, as discussed in \S\S 3.1 and 3.2 the observed
ellipticity gradient and isophote twist suggest that this
assumption may not be correct.  The key improvement in the models
is the addition of internal structure in the lens galaxy's mass
comparable to that seen in its light.  While we do not expect the
mass to trace the light exactly, it should be given the same
freedoms in order to avoid oversimplifying the models.

Bernstein \& Fischer (1999) did add freedom to the galaxy's radial
and angular structure by using independent power laws in different
radial zones, but this led to unphysical models with density
discontinuities that cannot match normal rotation curves.  We
introduce models that allow similar radial and angular freedom
while keeping the density smooth. We start with the pseudo-Jaffe
ellipsoid, whose projected density distribution has elliptical
symmetry and a profile that is roughly flat inside a core radius
$s$, falls as $\Sigma \propto R^{-1}$ out to a cut-off radius $a$
(yielding a rotation curve that is approximately flat), and then
falls as $\Sigma \propto R^{-3}$ to maintain a finite mass; a
detailed definition is given in Appendix B.  We then construct
``double pseudo-Jaffe'' models comprising two concentric
pseudo-Jaffe components with different scale lengths, ellipticities,
and orientations.  We fix the model galaxy to its observed position,
which leaves 10 galaxy parameters: a mass parameter ($b_i$),
ellipticity ($e_i$), orientation angle ($\PA_i$), core radius
($s_i$), and cut-off radius ($a_i$) for each component ($i=1,2$).

The double pseudo-Jaffe model provides a great deal of freedom in
the lens galaxy. First, it is essentially a smooth generalization
of both the power law and the FGS models. When the inner
pseudo-Jaffe component is compact, it mimics the point mass in FGS
models while the outer component is similar to the King model (see
Appendix B); and when the cut-off radius of the inner component
is comparable to the core radius of the outer component, the two
components combine to produce nearly a smooth $\al = 1$ power law.
Second, with this model we can mimic the observed internal structure
of the lens galaxy: by adjusting the ellipticities and orientations
we can produce a smooth ellipticity gradient and isophote twist.
An important difference from the broken power law models of
Bernstein \& Fischer (1999) is that the double pseudo-Jaffe models
have smooth density profiles. We discuss the physical properties
of sample double pseudo-Jaffe models in \S 5.3.

\subsection{Cluster models}

Previous models have used several different methods for including
the cluster's contribution to the lensing potential. The simplest
approach is to expand the cluster potential in a Taylor series and
keep only the lowest significant (2nd order) term, which describes
the tidal shear produced by the cluster. However, this approximation
is thought to be poor for Q~0957+561 because the cluster is close
to the lens and the VLBI errorbars are small, so the 3rd order
terms are larger than the errorbars. Barkana et al.\ (1999) and
Chae (1999) included all the higher order terms implicitly by
introducing a mass distribution to represent the cluster, but
degeneracies related to the cluster shape and density profile
forced them to make assumptions about the cluster properties.
Kochanek (1991) and Bernstein \& Fischer (1999) instead used a
Taylor series with general 3rd order terms,
\begin{equation}
  \phi_{clus} = {1\over2}\,\kc\,r^2
    + {1\over2}\,\g\,r^2 \cos 2(\th-\tg)
    + {1\over4}\,\s \,r^3 \sin  (\th-\ts)
    - {1\over6}\,\dd\,r^3 \sin 3(\th-\td)\,.
  \label{eq:phiclus}
\end{equation}
The 2nd order $\kc$ term produces the mass sheet degeneracy,
so it is usually omitted from the lens models and constrained
independently (see \S\S 3.1 and 3.2). The 2nd order $\g$ term
represents the tidal shear from the cluster. The 3rd order $\s$ and
$\dd$ terms arise from the gradient of the cluster density and the
$m=3$ component of the cluster mass (in a frame centered on the
lens galaxy; see Bernstein \& Fischer 1999; Keeton 2000). The three
direction angles $(\tg,\ts,\td)$ are written here as position
angles (i.e.\ measured East of North).

We follow Bernstein \& Fischer (1999) and use the 3rd order Taylor
series (omitting the $\kc$ term).  Although we simply fit for the
Taylor series parameters, it is important to understand how they
relate to physical properties of the cluster and what their
reasonable ranges are.  In popular cluster mass models the cluster
amplitudes $(\g,\s,\dd)$ and direction angles $(\tg,\ts,\td)$ have
the following properties (Keeton 2000):
\begin{itemize}
\item Circularly symmetric mass distribution: the three angles
  all point to the center of the cluster.
\item Singular isothermal ellipsoid: the shear angle $\tg$ points
  to the cluster center and the shear amplitude equals the
  convergence from the cluster, $\g = \kc$. Both of these results
  hold for all positions and all values of the axis ratio $q$.
  The gradient amplitude $\s$ is bounded by $1 \le {\s r_0 \over
  \kc} \le {1+q^2 \over 2q}$, where $r_0$ is the distance from
  the lens galaxy to the cluster.
\item Softened isothermal ellipsoid: Outside of the core the
  results for the singular isothermal ellipsoid are approximately
  true. For example, for clusters with an axis ratio $q>0.6$ the
  shear angle $\tg$ points to within $10\arcdeg$ of the cluster
  center for $r_0 \gtrsim 2s$, where $s$ is the core radius. The
  cluster has $\kc \ge \g$, but $\kc > 2\g$ only for $r_0 \lesssim 3s$.
\item An ellipsoid with the ``universal'' dark matter profile of
  Navarro, Frenk \& White (1996): For clusters with an axis ratio
  $q>0.6$, the shear angle $\tg$ points to within $4\arcdeg$ of the
  cluster center for $r_0 \gtrsim 0.5 r_s$, where $r_s$ is the scale
  radius in the ``universal'' profile. The cluster generally has
  $\kc \gtrsim \g$ for $r_0 \gtrsim r_s$ and $\kc \lesssim \g$ for
  $r_0 \lesssim r_s$.
\end{itemize}
We let the shear and gradient parameters $(\g,\tg,\s,\ts)$ vary
freely, but we keep these general relations in mind when
interpreting the results. To avoid an explosion of parameters, we
follow Bernstein \& Fischer (1999) and use a ``restricted'' 3rd
order cluster with $\td=\ts$ and $\dd=-3\s/2$ (as for a singular
isothermal sphere). This model is useful because it requires only
four parameters for the cluster, and it can provide
model-independent evidence for a non-circular cluster if $\ts \ne
\tg$. We also experimented with models using a singular isothermal
mass distribution for the cluster, and we discuss these models
briefly but do not quote detailed results (\S 5).

\subsection{Constraining the new models}

Our models have 10 galaxy and 4 cluster parameters, but fortunately
we need not examine all of them explicitly. First, the general
features of the lens allow us to fix two scale radii. The lack of
a central image requires that the galaxy be nearly singular, so we
set the core radius of the inner pseudo-Jaffe component ($s_1$) to
zero. Also, previous lens models suggest that the galaxy needs to
have mass extending at least to the distant A image ($5\arcsec$
from the galaxy), so the cut-off radius of the outer pseudo-Jaffe
component ($a_2$) should be larger than $5\arcsec$. The model
should not be very sensitive to any particular value larger than
this, so we fix $a_2 = 30\arcsec$ (similar to Bernstein \& Fischer
1999).

Second, we can use linear techniques for the two galaxy mass
parameters ($b_1$ and $b_2$) and the two cluster amplitudes ($\g$
and $\s$) because they enter the potential as simple multiplicative
factors. (This is a variant of solving for the shear parameters
in eq.~\ref{eq:shr}.) If we take the strong VLBI constraints on
the quasar core and jet positions to be exact (see Kochanek 1991;
Bernstein \& Fischer 1999), they lead to four linear constraint
equations:
\begin{eqnarray}
  \mbox{cores}:\qquad \xx_{\rm A1} - \nabla \phi(\xx_{\rm A1})
    &=& \xx_{\rm B1} - \nabla \phi(\xx_{\rm B1}) \\
  \mbox{jets}:\qquad  \xx_{\rm A5} - \nabla \phi(\xx_{\rm A5})
    &=& \xx_{\rm B5} - \nabla \phi(\xx_{\rm B5})
\end{eqnarray}
where A$_1$ and B$_1$ denote the two quasar cores while A$_5$ and
B$_5$ denote the brightest jet components; the positions are given
by Barkana et al.\ (1999). The four constraint equations can be
solved explicitly to write the four linear parameters as functions
of the remaining (non-linear) parameters. This technique ensures
that the VLBI constraints are fitted exactly, while reducing by
four the number of parameters that must be examined directly.

We are left with 8 explicit parameters: the ellipticities $e_1$ and
$e_2$, orientation angles $\PA_1$ and $\PA_2$, and scale lengths
$a_1$ and $s_2$ of the galaxy; and the direction angles $\tg$ and
$\ts$ of the cluster. We evaluate models in this parameter space
using the remaining constraints: the flux ratios of the quasar
cores and jets and the positions of the optical Blobs and Knots
(taken from Bernstein \& Fischer 1999), and the structure of the
arcs (see \S 3.3). Techniques and results for our new models are
discussed in \S 5.

\section{The failure of existing lens models}

Previous models of Q~0957+561 (summarized in Table 3) posited that
the VLBI observations and the optical Blobs and Knots provided
strong enough constraints to break the degeneracy between the
lens galaxy shape and the cluster shear.  As we discussed in \S 3.1,
however, it is dangerous to use smooth circular or elliptical lens
models with high-precision constraints, because the oversimplified
models may be forced to converge on a best-fit solution with the
wrong global structure for the potential.  Until now we have
lacked the constraints to test this concern, although the fact
that previous models gave wildly difference shapes for the lens
galaxy certainly suggests that the models were not robust.

The host galaxy arcs finally provide an extensive set of constraints
for testing the previous models.  The first clue comes
from comparing the qualitative features of the arcs and the existing
models. Despite differences in details, most of the previous models
have a lensing potential with a strong tidal shear from the cluster
($\g \sim 0.1$--$0.4$). The galaxy dominates the potential for the
close image (B), but the strong shear means that the cluster
dominates the potential for the distant image (A). The distinction
is important because in general a round host galaxy produces an
image distorted tangentially relative to the mass that dominates
the potential. As a result, for a circular host galaxy most previous
models of Q~0957+561 would predict a B arc tangential to the galaxy
(as observed), and an A arc tangential to the cluster -- i.e.\
stretched {\it radially\/} relative to the lens galaxy, opposite
what is observed. These lens models can produce an A image tangential
to the galaxy only if the source galaxy is highly flattened and
oriented at just the right angle. In other words, we must either
invoke a very special source configuration or conclude that the
standard Q~0957+561 models have generic problems.

We can quantify these problems by using the arc modeling technique
described in \S 3.3 to predict the structure of the B arc for each
model and compare it to the observed B arc. (Recall that this
technique uses the observed A arc to construct the source, so
models always reproduce the A arc correctly and hence are evaluated
by examining the B arc.) Figures 2 and 3 show the observed and
predicted models arcs for previous power law models, while Figure 4
shows the arcs for FGS models; Figure 5 shows the intrinsic source
inferred for two of the models. Table 3 includes the quantitative
$\bar\ca$ estimates for the models.

All of these models are inconsistent with the structure of the B
arc, both visually and in terms of $\bar\ca$. Although the
predicted arcs differ among the various models, there are two
common features. First, in some of the models (notably those with a
circular lens galaxy) the predicted B arc has a pair of bright
ridges not seen in the observed B arc. Models that produce such
ridges are strongly inconsistent with the data; they have $\bar\ca$
statistics no better than 5.0 and as poor as 10.8. Second, the
models generally fail to predict the correct shape for the B arc at
low surface brightness levels. The predicted arcs have too much
curvature and do not match the northeast extension of the observed
arc. The failure to match even the rough shape (extent and
curvature) of the B arc illustrates the point explained above: if
the potential near the A image is dominated by the cluster shear,
reproducing the A arc requires a flattened source (e.g.\ the
Barkana/SPEMD model in Figure 5) that is distorted into a highly
curved B arc. The existing models cannot match the arcs better than
$\bar\ca=2.2$, which given the large number of arc constraints
means that these models are formally excluded at an extremely high
significance level.

The models arcs also illustrate a crucial limitation of the broken
power law models of Bernstein \& Fischer (1999), represented here
by the DM2+C2 model in Figure 3. These models use independent
elliptical power law models in different radial zones to mimic the
ellipticity and orientation variations in the observed lens galaxy.
The problem is that the density is discontinuous across the zone
boundary, which leads to a discontinuity in the predicted B arc.
The discontinuity does not affect the quantitative $\bar\ca$
statistic because it occurs outside our mask, but it is
qualitatively unacceptable. Combining this with the inability of
the broken power law models to match normal rotation curves
emphasizes that the models are physically unacceptable. The need
for a more physically reasonable way to include ellipticity and
orientation variations motivates our introduction of the double
pseudo-Jaffe models.

Although the previous models do not correctly reproduce the
distortions of the host galaxy, they do reveal a model-independent
qualitative feature of the lensing. Part of the quasar host galaxy
is doubly-imaged like the quasar, but a small region crosses the
lensing caustic and is quadruply-imaged (see Figure 5). Arc A
is a single distorted image of the host galaxy, but arc B is a
combination of the three remaining images of the quadruply-imaged
and the one remaining image of the doubly-imaged region. This
fact explains why the A and B arcs have such different geometries.
It also implies that with the current observations of the arcs,
most of the constraints come from the small region of the host
galaxy that is quadruply-imaged. A deeper image of the arcs
should reveal a complete Einstein ring that would strengthen
the constraints by using more of the host galaxy (see \S 6).

In summary, the host galaxy arcs in Q~0957+561 offer a strong new
probe of the global shape of the lensing potential, and they show
that existing lens models converged to the wrong potential.
Generically, the previous models have too much shear from the
cluster. The failure to fit the arcs is a common problem of models
that combine strong constraints from the sub-milli-arcsecond
structure of the quasar jets with oversimplified circular or
elliptical lens models. It rules out existing models of the
system, together with the bounds on $H_0$ drawn from them.

\section{Successful new models}

Two goals motivate our search for new models of Q~0957+561. First,
if we want to use the VLBI constraints with sub-milli-arcsecond
precision, we must be sensitive to details of the lens galaxy
structure such as its radially varying ellipticity and orientation.
We used double pseudo-Jaffe models (see \S 3.4) to incorporate such
structure in a smooth and physically reasonable way. Second, we
want to incorporate the powerful constraints from the host galaxy
arcs into the modeling process, instead of using them for {\it a
posteriori\/} tests as in \S 4. In this way we hope to break the
environment degeneracy in a robust way. When we began to examine
new models we found a range of solutions consistent with the data,
so we adopted Monte Carlo techniques to sample this range. In this
section we first describe the Monte Carlo techniques (\S 5.1), and
then discuss the models and their implications for breaking the
degeneracy between the galaxy and cluster (\S 5.2), the physical
properties of the lens galaxy and cluster (\S 5.3), and the Hubble
constant (\S 5.4).

\subsection{Monte Carlo techniques}

To explore our model space, we picked random values for the
non-linear model parameters, using restricted but
physically-motivated ranges:
\begin{itemize}
\item Galaxy ellipticity: There are no successful models with
  an outer ellipticity below 0.3. We believed models with an
  ellipticity larger than $0.7$ to be implausible. Hence we
  considered $0.3 \le e_2 \le 0.7$ for the outer ellipticity.
  In models with an ellipticity gradient the inner ellipticity
  can be small, so we considered $0 \le e_1 \le 0.7$.
\item Galaxy orientation: In the observed lens galaxy the PA
  varies from $\sim$40\arcdeg\ to $\sim$60\arcdeg\ (Bernstein
  et al.\ 1997). We expect the mass to be roughly aligned with
  the light (to within $\sim$10\arcdeg; Keeton et al.\ 1998),
  so we considered $30\arcdeg \le (\PA_1,\PA_2) \le 70\arcdeg$
  for the inner and outer position angles.
\item Galaxy scale lengths: We fixed the inner pseudo-Jaffe
  component to be singular ($s_1=0$) and the outer component
  to have a cut-off radius $a_2 = 30\arcsec$ (see \S 3.6).
  For the cut-off radius of the inner component ($a_1$) and
  the core radius of the outer component ($s_2$) we considered
  $0\farcs1 \le (a_1,s_2) \le 4\arcsec$.
\item Cluster direction angles: Galaxy counts and the weak
  lensing measurement suggest that the cluster has its mass
  concentration toward $\tc \sim 55\arcdeg$ (Fischer et al.\
  1997). For reasonable cluster models the shear angle $\tg$
  points roughly to the cluster center (see \S 3.5), so we
  considered shear angles in the range $30\arcdeg \le \tg \le
  70\arcdeg$. We were less restrictive with the gradient angle
  $\ts$ and required only that it be in the same quadrant as
  the cluster, $0\arcdeg \le \ts \le 90\arcdeg$.
\end{itemize}
Given values for these parameters, we fixed the four remaining
parameters ($b_1$, $b_2$, $\g$, $\s$) using the constraints from
the quasar core and jet positions (see \S 3.6).  Our limited
parameter ranges may omit some models that are formally consistent
with the data, but they span what we expect for physically
reasonable models.

We tabulated models that fit the data at the 95\% confidence
level.  With $N=(2,4,6)$ constraints this is equivalent to
$\chi^2 \le (5.99,9.49,12.59)$ (e.g.\ Press et al.\ 1992, \S\S 6.2
and 15.6).  We applied these thresholds to the constraints from
the quasars and the optical Blobs and Knots both separately and
jointly: we required $\cf \le 5.99$ for the flux ratios of the
quasar cores and jets (2 constraints), $\cbk \le 9.49$ for the
positions of the optical Blobs and Knots (4 constraints), and
$\cf + \cbk \le 12.59$ for the 6 joint constraints.  We did not
impose constraints from the less reliable Blob and Knot flux
ratios.  We considered different thresholds for the arc constraints,
as discussed below. Note that it was not useful to combine all
the constraints into a total $\chi^2/\mbox{DOF}$ because the number
of constraints from the arcs is so large.

The Monte Carlo technique is not especially efficient; we
examined $\sim\!10^7$ models and found of order 1000 models
consistent with the data. However, examining that many models is
not prohibitively time consuming. Most of the models can be ruled
out quickly because they fail the flux and Blob/Knot $\chi^2$
cuts. For the remaining models, we first computed a fast $\bar\ca$
using a 4 times undersampled arc map; only for promising models
did we compute $\bar\ca$ using the full arc map. The benefit of
this brute force approach is the ability to identify and sample
the wide range of models consistent with the data, and thereby
estimate the full range of possible $H_0$ values.

We first considered models in which the inner and outer components
of the model galaxy were fixed to have the same ellipticity and
orientation; we found $(3, 275, 832, 1575)$ models with $\bar\ca <
(1.2, 1.3, 1.4, 1.5)$, whose properties are summarized in Figure 6.
Next we allowed the inner and outer galaxy components to have
different shapes, mimicking ellipticity and orientation gradients.
For simplicity, we refer  to such models as having a ``twist,''
even though that term formally describes only an orientation
gradient.  We found $(25, 134, 286, 477)$ twist models with
$\bar\ca < (1.2, 1.3, 1.4, 1.5)$, whose properties are summarized
in Figure 7.  Table 4 gives parameters and Figure 8 shows the
predicted arcs for sample models of both types.  We note that
the number of successful models with a twist is smaller than the
number of successful models without a twist.  Introducing the
twist enlarges the parameter space without necessarily increasing
the space of successful models by the same amount; hence sampling
the same total number of models yields fewer acceptable models.
In the following sections we discuss models with $\bar\ca < 1.5$,
but our conclusions would not change substantially if we lowered
this threshold.

\subsection{Breaking the galaxy/cluster degeneracy}

The host galaxy arcs finally break the degeneracy between the
lens galaxy shape and the cluster shear in Q~0957+561.  First,
they constrain the lens galaxy shape.  Among the previous models,
the three with the best fits to the arcs have $62\arcdeg < \PA <
67\arcdeg$ (Table 3).  In the new models, the bounds on the (outer)
orientation of the lens galaxy are $55\arcdeg < \PA < 66\arcdeg$
in models without a twist (Figure 6d) and $52\arcdeg < \PA_2 <
68\arcdeg$ in models with a twist (Figure 7f).  These ranges are
consistent with the observed galaxy's orientation of about
$56\pm8\arcdeg$ (Bernstein et al.\ 1997).  In other words, the
arcs not only constrain the model galaxy's orientation, they
require that its mass distribution be at least roughly aligned
with the light distribution; such alignment is also seen in
other lens systems (Keeton et al.\ 1998).  Also, in the new
models the bounds on the (outer) ellipticity of the galaxy are
$0.38 < e < 0.61$ in models without a twist (Figure 6c) and
$0.35 < e_2 < 0.63$ in models with a twist (Figure 7e).  These
ranges are slightly higher than the ellipticity seen in the outer
parts of the galaxy (Bernstein et al.\ 1997), but there is no
{\it a priori\/} reason to expect the ellipticities of the dark
halo and the light to be similar. These constraints on the lens
galaxy shape span models with a wide range of lens galaxy mass
profiles (controlled by the scale radii $a_1$ and $s_2$; see
Figures 6b and 7b), ellipticity profiles (with or without a
twist), and cluster contributions. Thus we believe that they
are robust, and that the ability of the host galaxy arcs to
finally constrain the galaxy shape is the most important result
of our new models.

Second, the arcs place limits on the lensing contribution of
the cluster. The new models have moderate or even small cluster
shears (typically $\g \lesssim 0.05$; Figures 6e and 7g), which
contrasts with the strong shears in previous models.  A small
shear is required by the fact that both arcs are tangential to
the lens galaxy (see \S 4). Also, including the 3rd order cluster
term is important, as expected from the proximity of the cluster
to the lens (Kochanek 1991).  Most models have a cluster gradient
amplitude in the range $0.006 < \s < 0.016$ (Figures 6g and 7i),
so at the position of image A the 3rd order term is comparable
to or even larger than the 2nd order term (see eq.~\ref{eq:phiclus}).
Because the cluster amplitudes are small, the corresponding
direction angles $\tg$ and $\ts$ are poorly constrained (Figures
6f, 6h, 7h, and 7j). Nevertheless, the fact that they usually
differ implies that the cluster cannot be spherical (see \S 3.5).

Unfortunately, even the arcs cannot uniquely determine the
model.  First, the cluster angles can adjust to accommodate the
narrow but finite range of lens galaxies given above.  Second,
in models with a twist the inner component can take on all values
of ellipticity and orientation in the ranges we allowed (Figures
7c and 7d).  This is because the inner component is usually
compact, so like the point mass in FGS lens models it serves
mainly to correct a central mass deficit and the model is largely
insensitive to the distribution of that mass.  Both of these
degeneracies affect the potential enough to produce a disappointingly
wide range of Hubble constant values (see \S 5.4).

These conclusions are drawn from models using a Taylor series for
the cluster potential.  We also examined models treating the cluster
using an elongated isothermal model with an arbitrary position, axis
ratio, and orientation.\footnote{For technical reasons, the isothermal
cluster models we used had only approximate elliptical symmetry.}
The above conclusions about the galaxy properties still pertain,
with one modification: in models with a twist the distribution of
outer orientation angles gains a tail down to $\PA_2 \sim 40\arcdeg$.
The above conclusions about the cluster contributions also hold,
except that the new bounds on the shear amplitude are $0.03 < \g <
0.11$.  Since an isothermal cluster has $\g \approx \kc$, it is hard
to obtain a small shear without essentially eliminating the cluster.

\subsection{Properties of the galaxy and cluster}

We introduced the double pseudo-Jaffe models in order to smoothly
include ellipticity and orientation variations in the lens galaxy,
but we must ask whether the models are physically plausible. Figure
9 shows the estimated density and circular velocity profiles for
the models in Table 4. These profiles are only estimates because
they require the full 3-d mass distribution, while lens models
give only the projected distribution.  We assumed that the
intrinsic distributions are oblate spheroids viewed edge-on, and
we computed the density and circular velocity profiles in the
equatorial plane, neglecting any twist.  The high inferred circular
velocities ($v_c \sim 600$ \kms) are misleading, because our models
omit any convergence from the cluster and $v_c \propto (1-\kc)^{1/2}$,
and because converting to an observed velocity dispersion involves
systematic uncertainties in the stellar dynamics (see Romanowsky
\& Kochanek 1999).  The qualitative features, however, are plausible.
The inner and outer galaxy components generally combine to produce
a rotation curve that is approximately flat, although in several
models it turns up at small radii because the inner component of
the pseudo-Jaffe models is compact.  This effect, which also
appears in FGS models because of the point mass, occurs because
lensing constrains only the total mass enclosed by the B image
($r_B = 1\farcs0$), and if that mass is compact then it produces
a rising rotation curve.  We conclude that the fitted double
pseudo-Jaffe models are fairly reasonable.

As for the cluster, recall that the host galaxy arcs require the
shear to be small.  At the same time, two independent estimates of
the cluster mass suggest that the convergence from the cluster is
$\kc \sim 0.2$.  First, stellar dynamical modeling of the lens
galaxy constrains its mass such that explaining the image separation
requires a cluster convergence of $\kc = 0.20\pm0.12$ (Romanowsky
\& Kochanek 1999).  Second, Fischer et al.\ (1997) and Bernstein
\& Fischer (1999) used weak lensing to measure the total mass
inside an aperture centered on the lens galaxy. Subtracting the
mass of a model lens galaxy then gives a model-dependent estimate
of the remaining cluster mass (see eq.~\ref{eq:kc}).  With this
technique all of our models yield $0.20 \le \kc \le 0.23$.

These results are surprising because with popular cluster models
it is difficult to produce a shear that is much smaller than the
convergence (Keeton 2000).  Specifically, with a singular isothermal
ellipsoid, $\g=\kc$ for all cluster positions, orientations, and
axis ratios.  Introducing a core radius allows $\g \lesssim \kc$,
but the shear is substantially smaller than the convergence only
if the lens galaxy is near or within the cluster core.  If the
cluster halo has the ``universal'' or NFW density profile (Navarro
et al.\ 1996), $\g \lesssim \kc/2$ only inside $\sim\!0.5 r_s$,
where $r_s$ is the NFW scale length.  For the observed cluster at
redshift $z_l=0.36$ with a velocity dispersion of $\sim$700 \kms\
(Angonin-Willaime et al.\ 1994), the NFW scale length would be
$r_s \sim 100$ $h^{-1}$ kpc or $\sim\!30\arcsec$ (e.g.\ Navarro
et al.\ 1996).  Combining these results with the estimates of the
cluster contribution ($\g \lesssim 0.1$ or even $\lesssim 0.05$,
and $\kc \sim 0.2$) suggests two possible conclusions about the
cluster.  On the one hand, if the cluster is ellipsoidal it must
have a core or scale radius large enough to encompass the galaxy.
Having the lens galaxy be in the cluster core would not be too
surprising since it is the brightest galaxy in the cluster.  On
the other hand, the assumption of an ellipsoidal model for the
cluster may not be correct, perhaps because of substructure.

\subsection{Implications for the Hubble constant}

Finally, we want to combine the lens models and the observed time
delay and use \refeq{tdel} to determine the Hubble constant $H_0$.
As discussed in \S 3.1, the inferred value of $H_0$ depends weakly
on the other cosmological parameters; we quote results assuming
$\Omega_0=1$ and $\Lambda_0=0$ and note that they would increase
by 5.8\% (4.5\%) for an $\Omega_0=0.3$ open (flat) cosmology.
Also, the mass sheet degeneracy implies $H_0 \propto (1-\kc)$ (see
eq.~\ref{eq:hk}), and $\kc$ cannot be constrained by lens models.
Thus if we write $H_0 = 100\,h\,\Hunits$, lensing directly measures
only the combination $h/(1-\kc)$, which is what we quote in Tables
3 and 4 and Figures 6 and 7.  We need an independent estimate of
$\kc$ in order to constrain $H_0$ itself.

We saw in \S 5.2 that the host galaxy arcs help break the
degeneracy between the lens galaxy shape and the cluster shear,
but they do not eliminate it entirely.  The remaining freedom in
the models is small in terms of the properties of the lens galaxy
but large in terms of the Hubble constant.  Models without a twist
yield $1.1 \lesssim h/(1-\kc) \lesssim 1.4$, with a tail down to
$h/(1-\kc) \approx 0.95$ (Figure 6). Allowing a twist in the lens
galaxy broadens the distribution down to $h/(1-\kc) \approx 0.85$
(Figure 7). In other words, within the lens models there is a
$\pm 25\%$ variation in inferred values for the Hubble constant;
this is independent of uncertainties in $\kc$. Most of this
variation is related to a strong correlation between the lens
galaxy ellipticity and the Hubble constant (Figures 6c, 7c, and
7e): flatter galaxies have deeper central potential wells when
normalized to produce the same images.

Although the model uncertainties limit our ability to use Q~0957+561
to constrain the Hubble constant, we might invert our thinking and
ask what properties of the lens model are required to be consistent
with local distance ladder determinations of $H_0$ (e.g.\ Mould et
al.\ 1999).  Assuming a cluster convergence $\kc \sim 0.2$ (see
\S 5.3), obtaining $H_0 \lesssim 80$ \Hunits\ requires a lens galaxy
that is relatively round in the center ($e_1 \lesssim 0.30$) and
moderately flattened in the outer parts ($0.35 \lesssim e_2
\lesssim 0.56$).  These bounds bolster the suggestion that the
lens galaxy must have a mass distribution with an ellipticity
gradient fairly similar to that seen in the light distribution.

\section{Conclusions}

We have detected large, distorted images of the quasar host
galaxy in the gravitational lens Q~0957+561.  In the H band
(rest frame R), the host galaxy appears as two long
($\sim$5\arcsec) arcs stretched tangentially relative to the
lens galaxy.  Previously published models of Q~0957+561 fail
to predict the correct shape for the host galaxy arcs.  This
failure rules out those models and any conclusions about the
value of the Hubble constant drawn from them.

The problem with the previous models is that they oversimplified
the mass distribution for the lens galaxy, so they had to
adjust the large-scale features of the model (the galaxy shape
and the cluster shear) in order to fit small-scale constraints
(the VLBI jets, with $\sim$0.1 milli-arcsecond errorbars).
Without strong constraints on the global shape of the lensing
potential, each class of models happily converged to a best-fit
solution with the wrong global structure.  The failure of these
models presents two important lessons (also see Kochanek 1991;
Bernstein et al.\ 1993; Mao \& Schneider 1998; Bernstein \&
Fischer 1999).  First, when the constraints are extremely
precise it is necessary to include the full complexity of the
structure of the lens galaxy.  Second, it is important to
carefully explore the full range of possible models before
leaping to conclusions about the value of the Hubble constant
and its uncertainties.

Two improvements are crucial to finding better models.  First,
the host galaxy arcs finally provide enough constraints to
separately determine both the lens galaxy shape and the cluster
shear, and they must be incorporated into the modeling process.
Second, models for the lens galaxy must allow internal structure
similar to that seen in the observed galaxy, which for Q~0957+561
means a radially varying ellipticity and orientation.  Although
the mass need not exactly trace the light, it must be given the
same freedoms lest we oversimplify the models.  We introduced a
new class of lens models, the double pseudo-Jaffe models, that
smoothly incorporate an ellipticity gradient and isophote twist.

Our models lead to several new conclusions about the system. First,
the shape of the lens galaxy's mass distribution must be
surprisingly similar to the shape of its luminosity distribution.
The mass distribution must be moderately flattened and roughly
aligned with the light distribution. Such alignment between the
mass and the light is also seen in other gravitational lenses
(Keeton et al.\ 1998), but was not seen in previous models of
Q~0957+561. Second, the shear from the cluster must be small ($\g
\lesssim 0.1$), in marked contrast with the moderate or strong
shears seen in previous models ($\g \sim 0.1$--$0.4$). The small
shear combined with an estimated convergence $\kc \sim 0.2$ implies
that the cluster potential must be approximately centered on the
lens galaxy. Unfortunately, the current observations of the host
galaxy cannot fully determine the lens model. There are still
freedoms related to several direction angles that describe the
cluster and to the amplitudes of the ellipticity and orientation
gradients, and they leave a 25\% uncertainty in the inferred value
of the Hubble constant.

However, there are two promising prospects for further improving
the constraints on this systems. First, the substantial progress we
have found is based on a 2800-second H band image. A deeper image
with a refurbished NICMOS camera should show a complete Einstein
ring image. Since an Einstein ring probes the potential all the way
around the lens galaxy, it is extremely useful for determining both
the lens model and the intrinsic shape of the host galaxy (see
Keeton, Kochanek \& McLeod 2000). Filling in the gaps in the
Q~0957+561 ring should eliminate the remaining uncertainties in the
models. Second, X-rays from the cluster gas have been detected
(Chartas et al.\ 1998), but the image resolution was poor and the
signal was dominated by X-rays from the quasar images. New
high-resolution X-ray observations to map the cluster gas would
constrain the cluster potential and help determine the cluster
angles that are still unknown (namely the angle to the cluster
center and the angle of the cluster's density gradient at the
lens). Combining deeper infrared and X-ray imaging would thus
dramatically improve the constraints on the models and allow
consistency checks of the cluster potential. Improved X-ray imaging
would also further improve the mass estimates of the cluster for
breaking the mass sheet degeneracy.

Distorted images of the host galaxy have now been observed in four
of the time delay lenses (Q~0957+561, PG~1115+080, B~1600+434, and
B~1608+656; see Impey et al.\ 1998; Kochanek et al.\ 1999). As we
have illustrated, images of the host galaxy are a powerful
constraint on models of the system and hence on the uncertainties
in the value of $H_0$ derived from the time delay measurements. To
date, host galaxies have been thought of mainly as pleasant bonuses
in relatively shallow images targeting the lens galaxies. Taking
full advantage of the host galaxies will require deeper images
focused on the host galaxies themselves. Such images hold great
promise for breaking common degeneracies in lens models and allowing
gravitational lensing to map in detail the mass distributions of
distant galaxies, to probe the potentials of lens galaxy
environments, and to determine a robust and independent measurement
of the Hubble constant at cosmological distances.

\acknowledgements

Acknowledgements: We thank Gary Bernstein for discussions and
comparisons with new optical data. Support for the CASTLES project
was provided by NASA through grant numbers GO-7495 and GO-7887 from
the Space Telescope Science Institute, which is operated by the
Association of Universities for Research in Astronomy, Inc., under
NASA contract NAS 5-26555.


\appendix

\section{Appendix: Comparing lens models}

Table 3 summarizes results from models of Q~0957+561 by Grogin
\& Narayan (1996), Barkana et al.\ (1999), Bernstein \& Fischer
(1999), and Chae (1999). The various authors reported results in
different ways. Sorting out the differences can be confusing, so we
converted all the results to a standard form and explain the
conversions here.

First, Bernstein \& Fischer (1999) and Chae (1999) quoted an
ellipticity parameter $\epsilon$ that is related to the axis ratio
$q$ by $q^2 = (1-\epsilon)/(1+\epsilon)$. We report the true
ellipticity $e = 1-q$. (Bernstein \& Fischer called their
ellipticity parameter $e$, but it played the same role as
$\epsilon$ here.)

Second, different authors quoted results for the Hubble constant in
different ways. In general, lens models can determine only the
combination $h/(1-\kc)$ where $h = H_0/(100\,\Hunits)$ and $\kc =
\Sc/\Scrit$ is the surface mass density of the cluster (in critical
units) at the position of the lens. Models that use a multipole
expansion for the cluster (an external shear or 3rd order cluster,
see eq.~\ref{eq:phiclus}) offer no constraints on $\kc$. Models
that use an actual mass distribution for the cluster do predict
$\kc$: an isothermal sphere that has core radius $s$ and is located
a distance $d$ from the lens produces $\kc = b/(2\xi)$ and shear
$\gc = (b d^2)/[2 \xi (s+\xi)^2]$, where $\xi = \sqrt{s^2+d^2}$ and
$b$ is a mass parameter such that $M(R) = \pi \Scrit\, b\,
(\sqrt{s^2+R^2}-s)$. However, the $\kc$ prediction is not unique
because varying the cluster's shape and profile can change $\kc$
(and $\gc$) without changing the goodness of fit (see Chae 1999 for
examples). Hence we describe these cluster models in terms of an
equivalent external shear with magnitude $\geff = \gc/(1-\kc)$ (see
Grogin \& Narayan 1996; Barkana et al.\ 1999). In Table 3 we quote
the magnitude of the external shear ($\g$, or $\geff$ in the
isothermal sphere cluster models) and the Hubble constant
combination $h/(1-\kc)$. We note that if the cluster is a singular
isothermal sphere, $(1-\kc) = (1+\geff)^{-1}$ (see Barkana et al.\
1999). The estimates of $h/(1-\kc)$ can be translated into actual
estimates for $H_0$ using the weak lensing measurement of the
cluster mass (see \S 3.2).

Third, different authors used slightly different definitions of
$\chi^2$. We did {\it not\/} convert $\chi^2$ values; we used what
the authors reported.

Finally, we corrected several apparent typographical errors.
Barkana et al.\ (1999) claimed that their galaxy orientation angle
$\th$ was a position angle, but in fact the position angle was
$90\arcdeg-\th$ and this is what we quote. Bernstein \& Fischer
(1999) quoted angles measured from the positive $x$-axis. Their
formalism is self-consistent, except that their eq.~(16) needs a
minus sign in front of the $\g$ term. We report the Bernstein \&
Fischer (1999) angles as position angles. Finally, for the
Bernstein \& Fischer (1999) 3rd order cluster models there appears
to be a difference of a factor of 2 between the amplitude of 3rd
order cluster term as written in their eq.~(16) and as reported in
their Table 3. We quote values consistent with their eq.~(16).

\section{Appendix: The pseudo-Jaffe ellipsoid}

A standard Jaffe (1983) model has a 3-dimensional density
distribution $\rho \propto m^{-2}\,(m+a)^{-2}$ where $m$ is an
ellipsoidal coordinate and $a$ is the break radius. For lensing it
is more convenient to use a modified density distribution $\rho
\propto (m^2+s^2)^{-1}\, (m^2+a^2)^{-1}$ where $a$ is again the
break radius, and we have added a core radius $s<a$. For this model
the projected surface mass density, in units of the critical
surface density for lensing, is
\begin{equation}
  {\Sigma \over \Scrit} = {b \over 2}\,\left[ \left(m^2+s^2\right)^{-1/2}
    - \left(m^2+a^2\right)^{-1/2} \right] , \label{eq:density}
\end{equation}
where the mass normalization parameter $b$ is chosen to match the
lensing critical radius in the limit of a singular isothermal
sphere ($s \to 0$, $a \to \infty$, $q\to 1$).
Eq.~(\ref{eq:density}) defines what we call the pseudo-Jaffe
ellipsoid. Its projected surface density is roughly constant for $R
\lesssim s$, falls as $R^{-1}$ for $s \lesssim R \lesssim a$, and
falls as $R^{-3}$ for $R \gtrsim a$; its total mass is $M = \pi
\Scrit\, q\, b\, (a-s)$. The ellipsoid coordinate $m$ can be written
in terms of the projected axis ratio $q$ and the position angle
$\PA$ (measured East of North) as
\begin{equation}
  m^2 = {R^2 \over 2q^2}\,\left[ \left(1+q^2\right)
    + \left(1-q^2\right)\cos2\left(\th-\PA\right) \right] .
\end{equation}
The pseudo-Jaffe lens model is easy to compute because it is
written as the difference of two softened isothermal ellipsoids,
whose analytic lensing properties are known (Kassiola \& Kovner
1993; Kormann, Schneider \& Bartelmann 1994; Keeton \& Kochanek
1998). This model has been used previously by de Zeeuw \& Pfenniger
(1988), Brainerd, Blandford \& Smail (1996), and Keeton \& Kochanek
(1998).

In the limit of a singular model ($s=0$), the pseudo-Jaffe model
is an example of a general class of ``cuspy'' lens models with
$\rho \propto m^{-\gamma} (m^2+a^2)^{(\gamma-4)/2}$. These models
are a more realistic family for real galaxies than the softened
power law models, and are discussed by Mu\~noz, Kochanek \& Keeton
(2000).

The pseudo-Jaffe model is related to the King model used in FGS
lens models. The standard approach is to approximate the King model
with a combination of isothermal models (Young et al.\ 1980),
\begin{equation}
  {\Sigma \over \Scrit} =
    { 2.12 b \over \sqrt{m^2 + 0.75 r_s^2} } -
    { 1.75 b \over \sqrt{m^2 + 2.99 r_s^2} } \, .
\end{equation}
The King model has a single scale radius $r_s$, while the
pseudo-Jaffe model has independent core and break radii. Also, in
the King model the coefficients of the two terms differ, while in
the pseudo-Jaffe model they are the same.



\clearpage
\centerline{\psfig{file=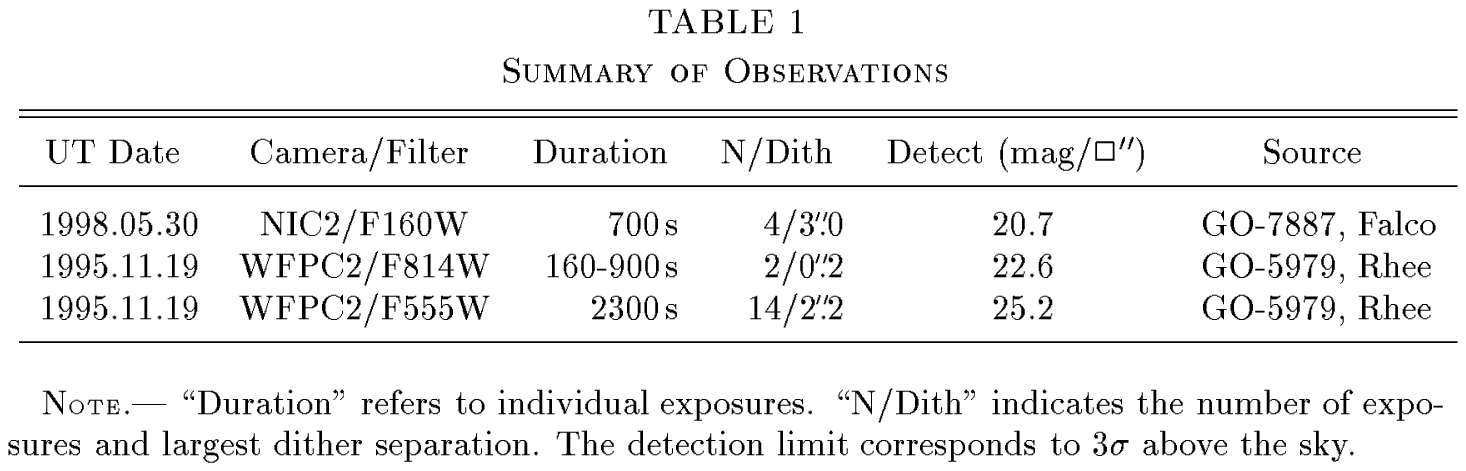}}
\centerline{\psfig{file=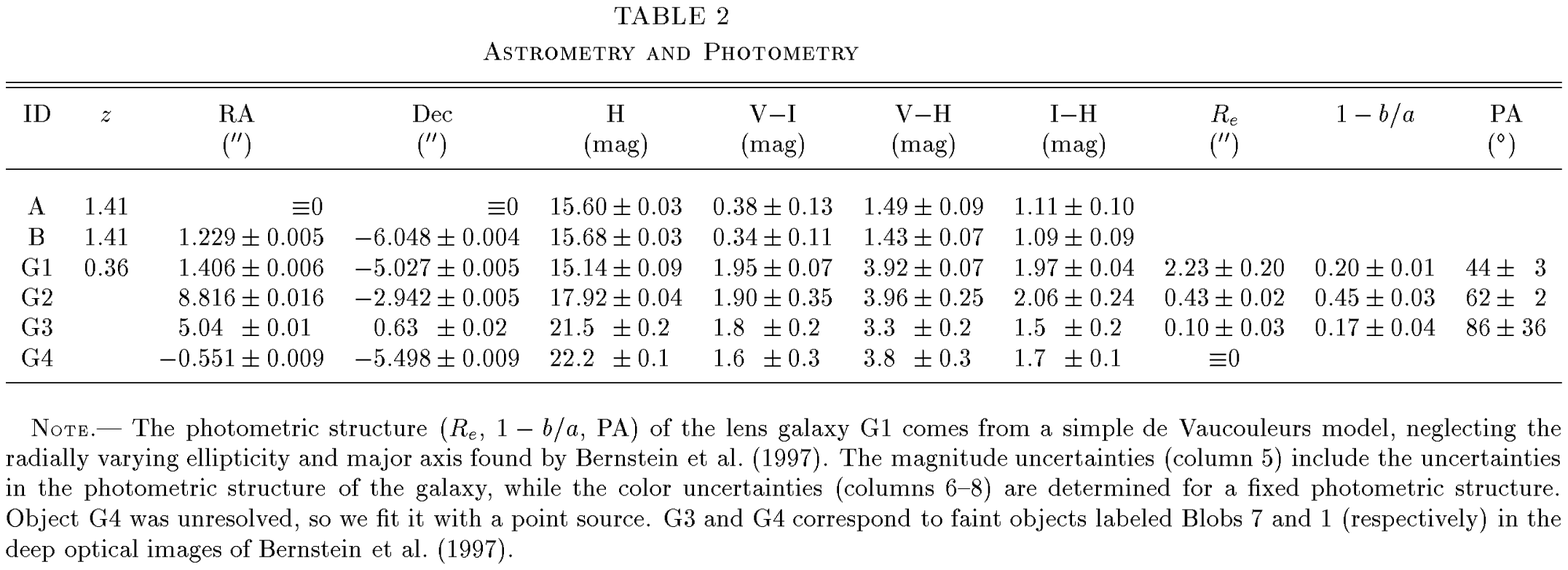,height=8.0in,angle=90}}
\centerline{\psfig{file=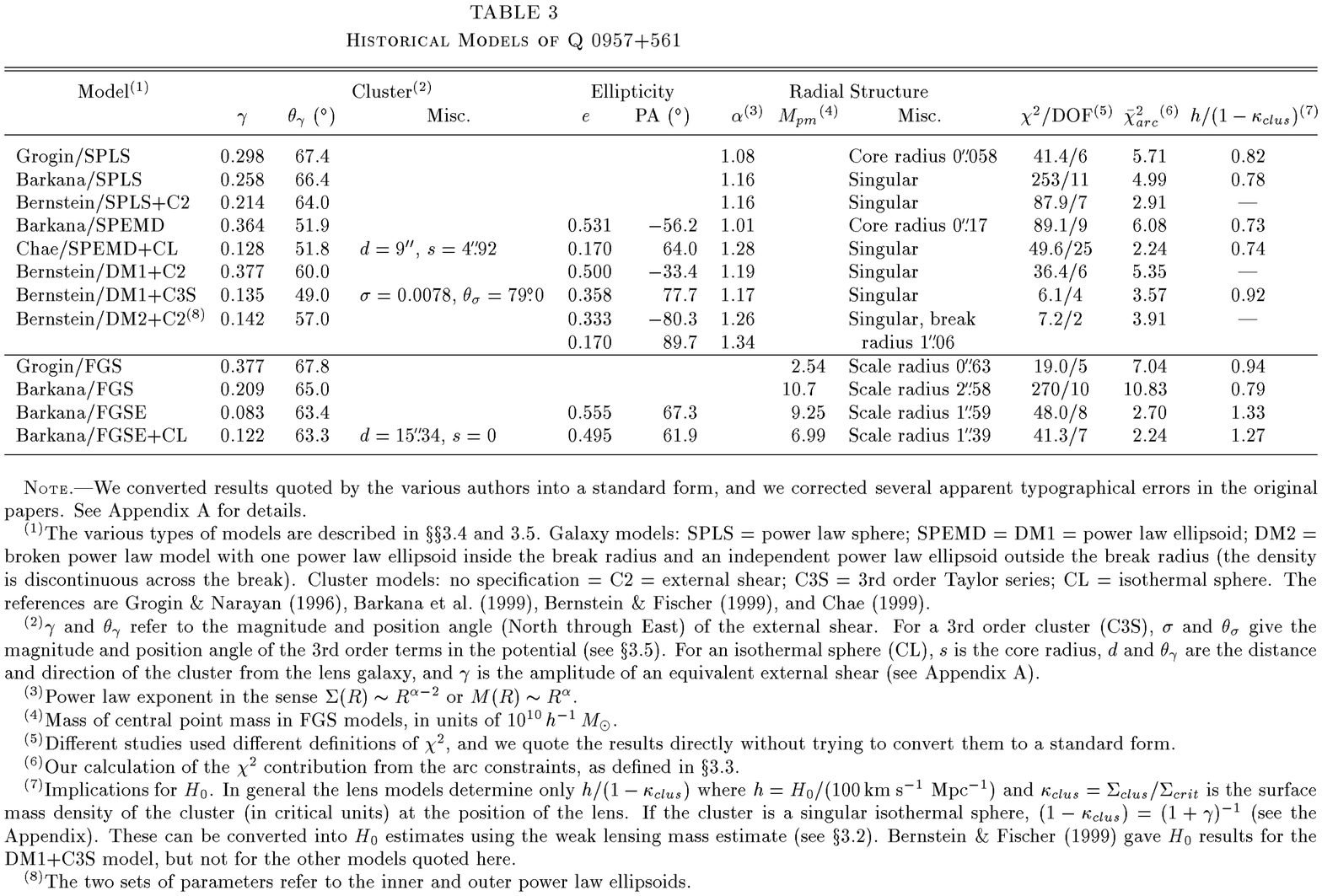,height=8.0in,angle=90}}
\centerline{\psfig{file=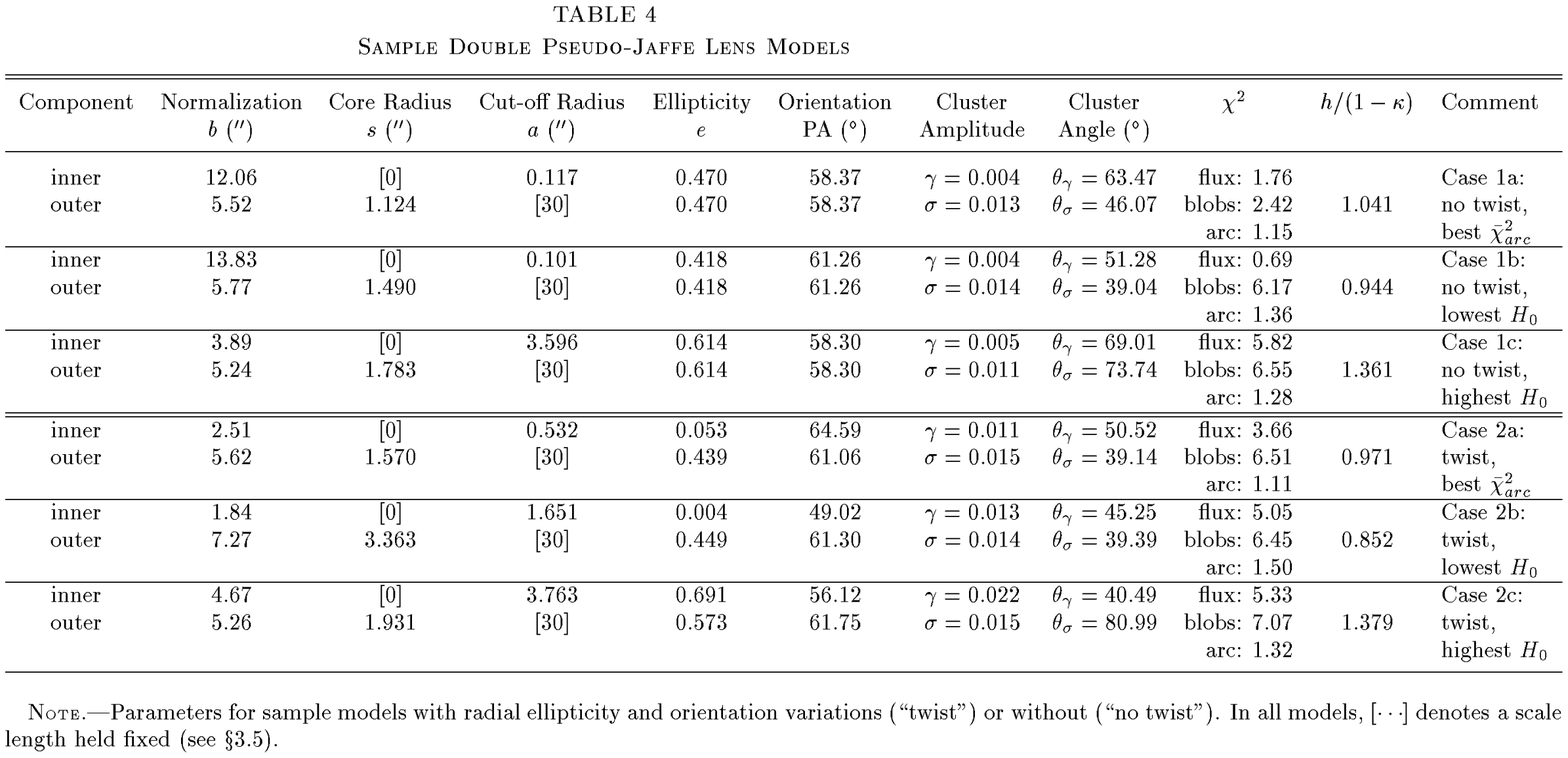,height=8.0in,angle=90}}


\begin{figure}
\centerline{\map{fig1a.ps}}
\centerline{\map{fig1b.ps}
            \map{fig1c.ps}}
\caption{
(a) The raw F160W (H band) image of the Q~0957+561 system, using a
logarithmic grayscale.  The lens galaxy (G1) and quasar images (A
and B) are labeled.  The tickmarks on the frame are spaced by
1\arcsec.
(b) The residual image after subtracting the two quasars and the
lens galaxy, plotted with a linear grayscale.  The holes in the bright
regions of the arcs are due to imperfect subtraction of the quasar
images.
(c) The residual image convolved with a $\sim$0\farcs076 (1 pixel)
FWHM Gaussian to enhance the visibility of low surface brightness
features.  The dotted contours are drawn at $2\sigma$ above the
sky.  The rectangular mask around the A arc shows the region used
to construct the unlensed source, while the polygonal mask around the
B arc shows the region used to compute $\ca$ (see \S 3.3).
}
\end{figure}

\begin{figure}
\centerline{\map{fig2a.ps}
            \map{fig2b.ps}}
\centerline{\map{fig2c.ps}
            \map{fig2d.ps}}
\caption{
The observed (grayscale) and predicted (contours) arc structures
for power law galaxy models. The grayscale is linear, and the solid
contours are drawn at $2,6,10,\ldots \times \sigma$ above the sky.
The dotted contours are drawn for the observed image at $2\sigma$.
The tickmarks on the frames are again spaced by 1\arcsec. We show
the Grogin \& Narayan (1996) SPLS model (top left), the Barkana et
al.\ (1999) SPLS (top right) and SPEMD (bottom right) models, and
the Chae (1999) SPEMD+CL model (bottom left). The galaxy is
circular in the SPLS model and elliptical in the SPEMD models. The
cluster is treated as an external shear in all models except for
Chae/SPEMD+CL, in which it is treated as a softened isothermal
sphere.
}
\end{figure}

\begin{figure}
\centerline{\map{fig3a.ps}
            \map{fig3b.ps}}
\centerline{\map{fig3c.ps}
            \map{fig3d.ps}}
\caption{
Power law galaxy models from Bernstein \& Fischer (1999). The
models are: SPLS+C2 (top left), DM1+C2 (top right), DM1+C3S (bottom
left), and DM2+C2 (bottom right). For the galaxy, SPLS indicates a
circular power law, DM1 indicates an elliptical power law, and DM2
indicates an elliptical broken power law. For the cluster, C2
refers to a 2nd order multipole expansion for the cluster (an
external shear), and C3S refers to a 3rd order expansion.
}
\end{figure}

\begin{figure}
\centerline{\map{fig4a.ps}
            \map{fig4b.ps}}
\centerline{\map{fig4c.ps}
            \map{fig4d.ps}}
\caption{
FGS-type models. We show the Grogin \& Narayan (1996) FGS model
(top left), as well as the Barkana et al.\ (1999) FGS (top right),
FGSE (bottom left), and FGSE+CL (bottom right) models. FGS refers
to a circular galaxy and FGSE refers to an elliptical galaxy;
both represent the cluster as an external shear. The FGSE+CL model
represents the cluster as a singular isothermal sphere.
}
\end{figure}

\begin{figure}
\centerline{\map{fig5a.ps}
            \map{fig5b.ps}}
\caption{
The intrinsic source structure inferred for two sample lens models.
The grayscale and contours are the same as in Figures 2--4, and
again the tickmarks on the frames are spaced by 1\arcsec. The heavy
curves show the lensing caustics. The cuspy diamond-shaped curve is
the tangential caustic, and the round curve in the Barkana/SPEMD
model is the radial caustic. In the Barkana/FGSE+CL model, the
radial caustic is formally at infinity because of the point mass. A
source inside the tangential caustic produces 4 images, between the
tangential and radial caustics produces 2 images, and outside the
radial caustic produces 1 image. The source for the Barkana/FGSE+CL
model is offset because of the deflection produced by the
isothermal cluster model.
}
\end{figure}

\begin{figure}
\centerline{\psfig{file=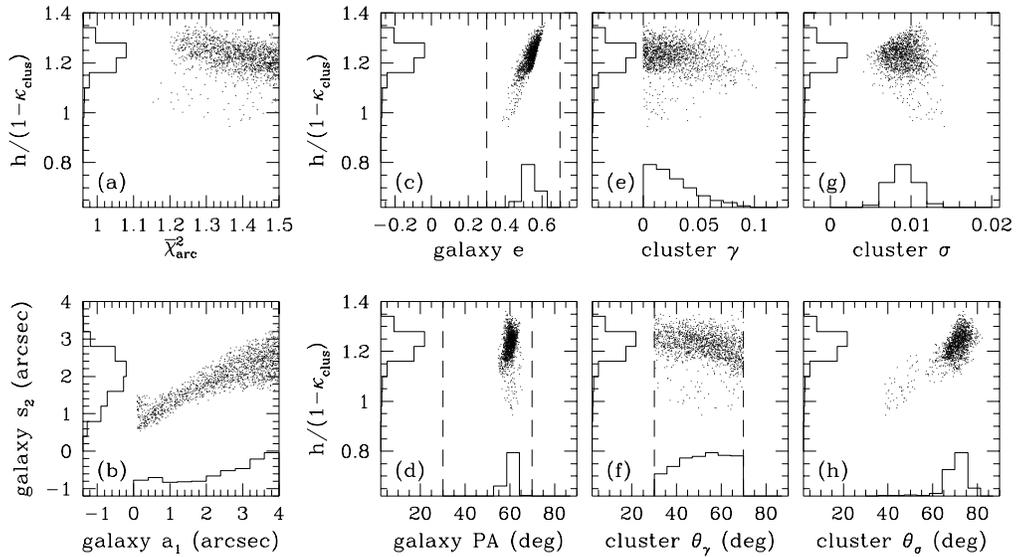,height=3.0in}}
\caption{
Results for models without radial ellipticity and orientation
variations. The points indicate the values for the parameters,
$\bar\ca$, and Hubble constant for individual models, while the
histograms show the corresponding distributions (normalized to have
the same peak value). We quote Hubble constant results in terms of
the quantity $h/(1-\kc)$ that is constrained by the lens models;
the Hubble constant histograms are the same in all panels. The
vertical dashed lines indicate the parameter ranges we examined,
except the range for $\ts$ is outside the figure ($0\arcdeg \le \ts
\le 90\arcdeg$). There were no explicit bounds placed on the
cluster amplitudes $\g$ and $\s$.
}
\end{figure}

\begin{figure}
\centerline{\psfig{file=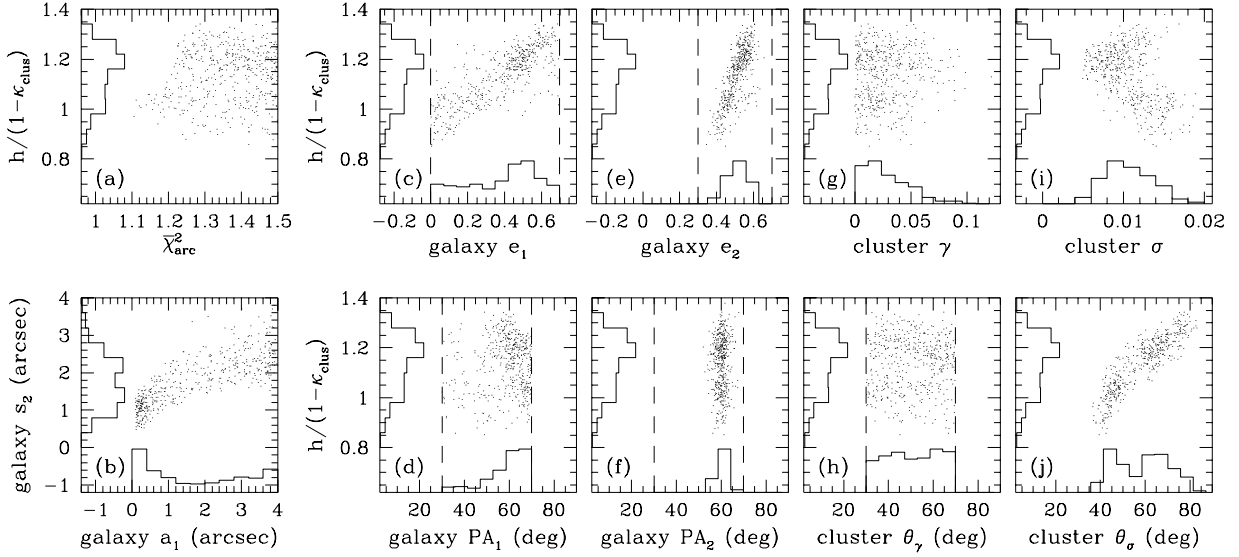,height=3.0in}}
\caption{
Similar to Figure 6 but for models with radial ellipticity and
orientation variations.
}
\end{figure}

\begin{figure}
\centerline{\map{fig8a.ps}
            \map{fig8b.ps}}
\centerline{\map{fig8c.ps}
            \map{fig8d.ps}}
\caption{
The observed and predicted arc structures for four of the new models
given in Table 4.  Cases 1a and 1b refer to the no-twist models that
have the best fit to the host galaxy arcs and the lowest value of
$H_0$, respectively.  Cases 2a and 2b refer to the twist models with
the same properties.
}
\end{figure}

\begin{figure}
\centerline{\psfig{file=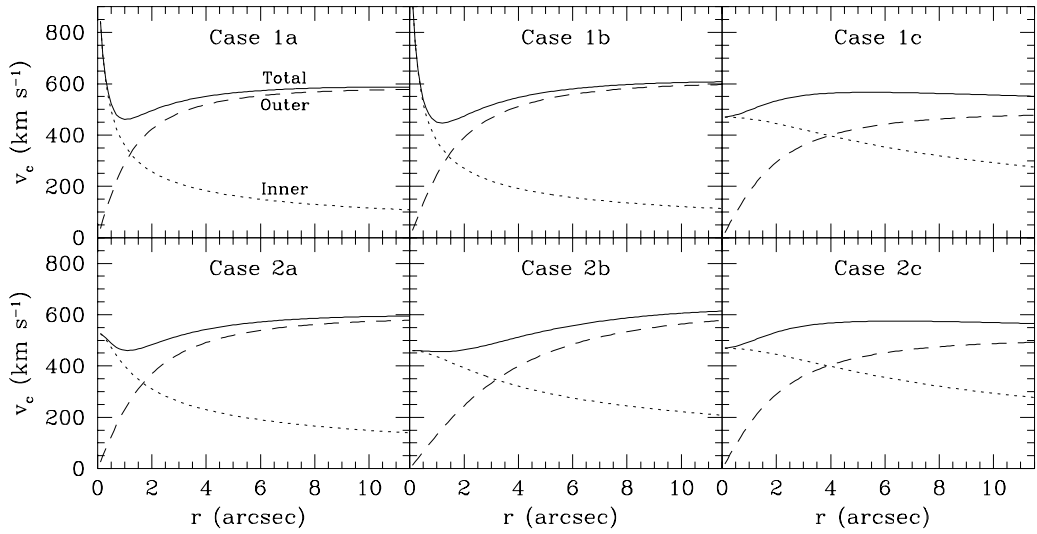,width=5.5in}}
\caption{
Estimates of the equatorial circular velocity profile for the
models given in Table 4, assuming that the 3-d mass distributions
are oblate spheroids viewed edge-on. The dotted and dashed curves
show the contributions from the inner and outer pseudo-Jaffe
components, respectively, and the solid curves show the total
rotation curves. For reference, the quasar images A and B are
5\farcs2 and 1\farcs0 from the center of the lens galaxy. Note
that the high inferred circular velocities ($v_c \sim 600$ \kms)
are misleading, as explained in the text.
}
\end{figure}

\end{document}